# A SCUBA-2 survey of FeLoBAL QSOs: Are FeLoBALs in a 'transition phase' between ULIRGs and QSOs?

Giulio Violino,[1] Kristen E. K. Coppin,[1] Jason A. Stevens,[1] Duncan Farrah,[2] James E. Geach,[1] Dave M. Alexander,[3] Ryan Hickox,[4] Daniel J.B. Smith,[1] Julie L. Wardlow[3,5]

[1] *Centre for Astrophysics Research, University of Hertfordshire, College Lane, Hatfield, AL10 9AB, UK.*
[2] *Department of Physics, Virginia Tech, Blacksburg, VA 24061, USA.*
[3] *Centre for Extragalactic Astronomy, Department of Physics, Durham University, South Road, Durham, DH13LE, UK.*
[4] *Department of Physics and Astronomy, Dartmouth College, 6127 Wilder Laboratory, Hanover, NH 03755, USA.*
[5] *Dark Cosmology Centre, Niels Bohr Institute, University of Copenhagen, Denmark.*

17 December 2015

**ABSTRACT**

It is thought that a class of broad absorption line (BAL) QSOs, characterised by Fe absorption features in their UV spectra (called 'FeLoBALs'), could mark a transition stage between the end of an obscured starburst event and a youthful QSO beginning to shed its dust cocoon, where Fe has been injected into the interstellar medium by the starburst. To test this hypothesis we have undertaken deep SCUBA-2 850 $\mu$m observations of a sample of 17 FeLoBAL QSOs with $0.89 \leqslant z \leqslant 2.78$ and $-23.31 \leqslant M_B \leqslant -28.50$ to directly detect an excess in the thermal emission of the dust which would probe enhanced star-formation activity. We find that FeLoBALs are not luminous sources in the submillimetre, none of them are individually detected at 850 $\mu$m, nor as a population through stacking ($F_s = 1.14 \pm 0.58$ mJy). Statistical and survival analyses reveal that FeLoBALs have sub-mm properties consistent with BAL and non-BAL QSOs with matched redshifts and magnitudes. An SED fitting analysis shows that the FIR emission is dominated by AGN activity, and a starburst component is required only in 6/17 sources of our sample; moreover the integrated total luminosity of 16/17 sources is $L \geqslant 10^{12} L_\odot$, high enough to classify FeLoBALs as infrared luminous.

In conclusion, we do not find any evidence in support of FeLoBAL QSOs being a transition population between a ULIRG and an unobscured QSO; in particular, FeLoBALs are not characterized by a cold starburst which would support this hypothesis.

## 1 INTRODUCTION

It has been established that every massive, local spheroid harbours a Supermassive Black Hole (SMBH, Kormendy et al. 1998; Kormendy et al. 1996) in its centre whose mass is proportional to that of its host bulge (e.g. Magorrian et al. 1998; Gebhardt et al. 2000), suggesting that there is a close connection between the central SMBHs and their surrounding galaxies. This hypothesis is supported by hydrodynamical galaxy formation simulations, which use feedback from active galactic nuclei (AGN) winds and jets to link the growth of the SMBH to that of its host (e.g. Di Matteo, Springel & Hernquist 2005; Hopkins et al. 2005; Bower et al. 2006). AGN feedback is thought to be responsible for quenching star formation in the host galaxy by heating up the interstellar medium and thinning out the reservoir of gas (Trouille et al. 2013; Yuma et al. 2013). This mechanism is a crucial component in the picture of Sanders et al. (1988b), where a starburst-dominated ultraluminous infrared galaxy (ULIRG), arising from a merger, evolves first into an obscured QSO and then into an unobscured QSO, where the AGN feedback had interrupted the previous phase of enhanced star formation activity. However, this is not the only scenario which can be invoked to explain the origin of QSOs, as a number of studies have showed that galaxy mergers and interactions only have a minimal impact on AGN activity (e.g. Sabater et al. 2015; Villforth et al. 2014), even for heavily obscured QSOs (Schawinski et al. 2012).

One direct way to test the model presented by Sanders et al. (1988b) observationally is to probe this evolutionary sequence at the 'transition stage' where the youthful QSOs are just beginning to shed their cocoons of gas and dust (e.g. Coppin et al. 2008; Simpson et al. 2012). However, given the combination of the implied high levels of obscuration and relatively short QSO lifetimes of $\sim 20-40$ Myr (Martini &Weinberg 2001; Goncalves et al. 2008) it has been difficult to select and confirm large samples of youthful QSOs.

There has been a long-running debate over the best way to find young QSOs (e.g. Sanders et al. 1988a), and much of the focus has been on the Broad Absorption Line (BAL) class of QSOs. BAL QSOs exhibit broad troughs ($\sim$ 2000–20,000 km s$^{-1}$ wide) in their UV and optical spectra arising from resonance-line absorption in gas with high outflowing velocities up to 66,000 km s$^{-1}$ (Lynds 1967; Weymann et al. 1991; Foltz et al. 1983; Hall et al. 2002; Reichard et al. 2003), comprising 26% of QSOs (Trump et al. 2006), and they come in 3 sub-types according to the visible absorption features: 1) High ionization BAL QSOs (HiBALs) show absorption in NV, SiIV and CIV; 2) Low ionization BAL QSOs (LoBALs) contain all of the HiBAL absorption features plus absorption in MgII and other low ionization species (Voit et al. 1993); and finally 3) the rarer FeLoBALs, which are LoBALs also exhibiting absorption from excited fine-structure levels or excited atomic terms of FeII or FeIII (e.g. Hazard et al. 1987; Becker et al. 1997, 2000). FeLoBALs comprise $\sim$0.3% of optically selected QSOs Trump et al. 2006, this fraction however increases by a factor of $\sim$10 when NIR and radio surveys are taken into account (e.g. Dai et al. 2012) due to the high level of dust obscuration which affects this class of quasars (Boroson & Meyers 1992, Allen et al. 2011). A scenario where BAL QSOs are young and are still surrounded by gas and dust from which the absorption features emanate was initially favoured, although others believed that the origin of the BAL features seen is more likely an orientation effect (where a BAL is a normal QSO seen along a line-of-sight which coincides with the outflowing gas; e.g. Elvis et al. 2000). Gallagher et al. (2007) found that the X-ray–to–far-infrared SEDs of HiBAL vs. non-BAL QSOs are indistinguishable – favouring the disk-wind paradigm with a typical radio-quiet QSO hosting a HiBAL region (e.g. Voit et al. 1993) in the AGN orientation unification scheme. Circumstantial evidence for LoBAL QSOs being young QSOs came from near-IR studies, which showed as LoBALs have redder optical continua, likely caused by dust absorption in the host galaxy and which could not be easily explained by orientation effects alone (e.g. Urrutia et al. 2009). However, sub-mm detection experiments show no relative difference between most BAL QSOs and non-BAL QSOs although the majority of both populations have a considerable number of upper limits (Lewis et al. 2003; Willott et al. 2003; Priddey et al. 2007, Cao Orjales et al. 2012).

More recently, the rarer FeLoBALs have emerged as the main contenders for young QSOs based on several lines of evidence 1) the only two systems at low-$z$ known to contain FeLoBALs are both ULIRGs (e.g. Farrah et al. 2005); 2) there is evidence for high-$z$ FeLoBALs in interacting systems (Hall et al. 2002; Gregg et al. 2002); and 3) the presence of large-scale winds in some FeLoBALs (de Kool et al. 2002) could provide a way for the emerging QSO to directly affect the star formation. Recently, Farrah et al. (2012) observed a large sample of 31 FeLoBALs with redshift $1 < z < 1.8$ in the mid-infrared with *Spitzer*, indicating that they are infrared-luminous.

By performing SED fitting using AGN and starburst templates, they claimed that star-formation is likely powering a relevant fraction of infrared output, although the AGN emission could be dominant. They also find an anti-correlation between the outflow absorption strength and the relative contribution to the infrared emission from a starburst component, which may indicate the disruptive effect of the AGN outflow on the obscured star formation. One way to investigate if the bolometric emission in these objects is dominated by star-formation is to detect the dusty star formation signature directly in the submillimetre, where the contamination from AGN emission is minimized. FeLoBALs would be easily detected with SCUBA-2 at 850 $\mu$m if they are forming stars at a prodigious rate (SFR $\sim$ 100's – 1000's M$_\odot$ yr$^{-1}$).

In this paper we investigate the validity of FeLoBALs being in a transition stage between a major starburst episode in ULIRG and an optically luminous QSO within the Sanders et al. (1988b) picture. Here we present a study of a sample of 17 FeLoBAL QSOs observed in the sub-mm to determine if that FeLoBALs have an enhanced dust content and star formation activity compared to other samples of QSOs (BAL and non-BAL), as expected if they are occurring at an earlier evolutionary state than normal QSOs.

The paper is organised as follows. The observations and data reduction are presented in Section 2. Section 3 describes the analysis and our main results. In section 4 the results are discussed in the general context of the far-infrared (far-IR) properties of BAL QSOs through a comparison with previous work. Finally we draw conclusions in Section 5. We adopt cosmological parameters from the *WMAP* fits in Spergel et al. (2003): $\Omega_\Lambda = 0.73$, $\Omega_{\rm m} = 0.27$, and $H_0 = 71 \,{\rm km \, s^{-1} \, Mpc^{-1}}$. All magnitudes are on the AB system unless otherwise stated.

## 2 DATA

### 2.1 Sample Selection and BAL properties

The parent sample comprises 138 FeLoBALs from Trump et al. (2006) (classified from the SDSS DR3; Schneider et al. 2005), as well as an incomplete list of 43 FeLoBALs identified in the BAL catalog of Gibson et al. (2009) (from SDSS DR5; Schneider et al. 2007). From these samples we select 9 FeLoBALs matched in redshift and magnitude with published 'benchmark' samples of submm-observed BAL and non-BAL QSOs in order to facilitate a direct comparison of the submm properties with these samples (Willott et al. 2003; Priddey et al. 2003; Priddey et al. 2007). In addition, we included 8 FeLoBALs from the Farrah et al. (2012) sample to allow us to explore the characteristics of FeLoBALs in a lower absolute magnitude regime and for which far-IR observations are available. Our final sample comprises 17 FeLoBALs (see Figure 1), with $-28.6 \leqslant {\rm M_B} \leqslant -23.3$, $0.89 \leqslant z \leqslant 2.78$, and balnicity indices $0 \leqslant {\rm BI} \leqslant 18000$ km s$^{-1}$ (see Table 2).

One source deserves particular attention: SDSSJ233646.20-010732.6. This QSO is certainly a FeLoBAL, however it belongs to a double system in which 4the separation between the two sources is $\leqslant 2"$, therefore its optical features, as well as its sub-mm emission, could be affected by its companion.

### 2.1.1 Definition of BAL QSO

The strength of the absorption features in the optical and UV spectra of BAL QSO is usually characterized through the balnicity index (BI). The BI represents the total velocity width over which the absorption exceeds a minimum value:

$$BI = \int_{v_0}^{v_1} (1 - \frac{f(-v)}{0.9}) C \, dv \qquad (1)$$

where $v$ is the velocity and $f(v)$ is the normalized flux at the velocity $v$. $C$ is a dimensionless quantity whose value is set to zero unless the observed absorption is at least 10% below the continuum for a certain velocity width, otherwise it is set to unity. Nonetheless this definition is far from being unambiguous, mainly because different authors choose different continuum levels and use different parametrizations for BI. We use as a first choice the modified BI by Gibson et al. (2009) where the selected species is the MgII and the values of $v_0$ and $v_1$ are respectively set to 0 and 25000 km s$^{-1}$ and the minimum velocity width to 2000 km s$^{-1}$.

For those objects in our sample which are not present in Gibson et al. (2009), we used the BI of Allen et al. (2011) where the sole difference is the value of $v_0$ which is set to 3000 km s$^{-1}$. Two of our FeLoBALs have quoted BI=0, this is due to the MgII absorption trough which is less than 10% below the continuum emission. However, these objects are classified as LoBALs once different species such as AlII are considered, or different balnicity measures are taken into account (i.e. Absorption Index, Hall et al. 2002; Trump et al. 2006) and we therefore decided to keep them in our sample. Moreover, for three FeLoBALs it is not possible to calculate the MgII BI. SDSSJ101108.89+515553.8 and SDSSJ114509.73+534158.1 have the MgII absorption line redshifted out of the spectral coverage, while SDSSJ233646.20-010732.6 belongs to a double system and its spectrum is likely a composite of the spectra of the two QSOs. As previously mentioned, the BAL characterization is not unique and the choice of a different species would bring about the same issues.

### 2.2 Observations and Data Reduction

Our sample was observed using the Submillimetre Common-User Bolometer Array 2(SCUBA-2, Holland et al. 2013) camera at the JCMT during two different runs: the first was a shared-risk observing run (S2SRO) in February 2010 and the second run was performed during routine operations in July 2012. The 225 GHz atmospheric zenith opacity was constantly monitored via the JCMT Water Vapour Monitor (WVM), and good conditions held throughout each night, with $0.05 < \tau < 0.08$ (where $\tau$ is the optical depth). Both 450 and 850 $\mu$m measurements were taken simultaneously. All observations were carried out in SCAN mode to make maps smaller than the field of view using the 'DAISY' pattern. Our primary goal was to integrate to a $1\sigma$ depth of $\simeq 2$ mJy at 850 $\mu$m, a comparable sensitivity to the benchmark comparison samples ($1\sigma \simeq 1.5$–3 mJy at 850 $\mu$m). The sensitivity calculator for both the S2SRO and the routine mode was used to estimate on-source integration times, of 16.5 and 46 min, respectively.

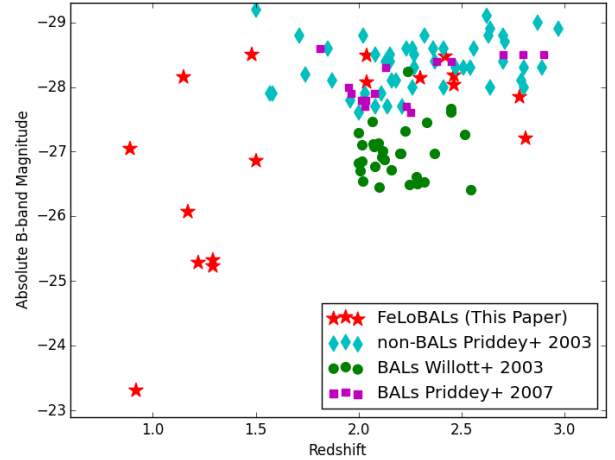

**Figure 1.** Overlap of our FeLoBAL QSO sample and comparison samples of 850 $\mu$m observed BAL and non-BAL QSOs. FeLoBALs cover a range of magnitudes $-28.6 \leqslant M_B \leqslant -23.3$ and redshifts $0.89 \leqslant z \leqslant 2.78$.

The SCUBA-2 data is reduced with the SMURF package (Jenness et al. 2011; Chapin et al. 2013), utilizing the Dynamic Iterative Map Maker (DIMM). After flat-fielding, the DIMM attempts to fit a model comprising (a) common mode signal (mainly atmospheric water and thermal emission), (b) astronomical signal including extinction correction and (c) a noise term. The DIMM iterates until convergence is met between the model and the data, or the fit no longer improves. After the bolometer time streams are mapped onto an astronomical grid, individual scans are co-added, weighting by inverse variance. Finally, a match-filter is applied using the *picard* routine SCUBA2-MATCHED-FILTER, which first removes any remaining large scale variation in the map still present after the main reduction steps above by smoothing the map with a $30''$ gaussian kernel, and then convolves the map with a model of the point spread function (see Dempsey et al. 2013). The final map is then calibrated using the flux conversion factors (FCFs) derived from observations of standard calibrators observed regularly since the start of SCUBA-2 operations. An additional 10% correction is added to account for flux lost during the matched-filtering step (see Geach et al. 2013). The absolute uncertainty on the flux calibration is around 15%, and we verify that the FCFs derived from calibrators observed during the project agree with the canonical values within the error bars. Since the SCUBA-2 camera observes simultaneously at 450 and 850$\mu$m, we report in Table 2 results derived from the data reduction at both wavelengths. However, due to the poor quality of the data at 450 $\mu$m, the following analysis is focussed primarily on the 850 $\mu$m measurements.

| Source | Tot. Int. Time | Observation dates |
|---|---|---|
| SDSSJ011117.34+142653.6 | 46 min. | 10/07/2012 |
| SDSSJ024254.66-072205.6 | 46 min. | 10/07/2012 |
| SDSSJ030000.57+004828.0 | 46 min. | 01/07/2012 |
| SDSSJ083817.00+295526.5 | 16.5 min. | 22/02/2010 |
| SDSSJ101108.89+515553.8 | 16.5 min. | 15,22/02/2010 |
| SDSSJ102850.31+511053.1 | 16.5 min. | 15,22/02/2010 |
| SDSSJ113424.64+323802.4 | 16.5 min. | 15,24,25/02/2010 |
| SDSSJ113734.06+024159.3 | 16.5 min. | 24/02/2010 |
| SDSSJ114509.73+534158.1 | 16 min. | 15,22/02/2010 |
| SDSSJ123549.95 +013252.6 | 46 min. | 01/07/2012 |
| SDSSJ131957.70+283311.1 | 16 min. | 15,22/02/2010 |
| SDSSJ135246.37+423923.5 | 16.5 min. | 15,25/02/2010 |
| SDSSJ142703.64+270940.3 | 46 min. | 01/07/2012 |
| SDSSJ155633.77+351757.3 | 46 min. | 01/07/2012 |
| SDSSJ210712.77+005439.4 | 46 min | 01/07/2012 |
| SDSSJ221511.93-004549.9 | 46 min. | 01/07/2012 |
| SDSSJ233646.20-010732.6 | 46 min. | 01/07/2012 |

**Table 1.** Details of the SCUBA-2 observations of FeLoBAL QSOs

| Source | $z$ | $M_B$ | $BI$ (km s$^{-1}$) | $F_{850}$ (mJy) | $<3\sigma$ 850 $\mu$m (mJy) | $<3\sigma$ 450 $\mu$m (mJy) |
|---|---|---|---|---|---|---|
| SDSSJ011117.34+142653.6 | 1.15 | −28.03 | 0* | 2.31 ± 2.30 | <9.21 | <45.95 |
| SDSSJ024254.66-072205.6 | 1.22 | −25.28 | 184.30 | 0.80 ± 2.37 | <7.91 | <51.82 |
| SDSSJ030000.57+004828.0 | 0.89 | −27.05 | 17372.4 | 4.65 ± 2.80 | <13.05 | <28.18 |
| SDSSJ083817.00+295526.5 | 2.04 | −28.08 | 0 | −3.72 ± 3.54 | <6.90 | <1071.80 |
| SDSSJ101108.89+515553.8 | 2.46 | −28.03 | − | 4.73 ± 2.62 | <8.85 | <277.34 |
| SDSSJ102850.31+511053.1 | 2.42 | −28.48 | 67.4 | 2.51 ± 2.34 | <9.53 | <182.94 |
| SDSSJ113424.64+323802.4 | 2.46 | −28.18 | 1371.7 | 0.84 ± 1.90 | <6.54 | <148.10 |
| SDSSJ113734.06+024159.3 | 2.78 | −27.12 | 14471.4* | 4.00 ± 3.95 | <15.85 | <40.69 |
| SDSSJ114509.73+534158.1 | 2.81 | −27.21 | − | 1.90 ± 2.02 | <7.96 | <417.01 |
| SDSSJ123549.95+013252.6 | 1.29 | −25.33 | 1313.9 | 0.11 ± 2.41 | <7.34 | <75.12 |
| SDSSJ131957.70+283311.1 | 2.04 | −28.49 | 232.0 | 0.40 ± 2.00 | <6.40 | <191.40 |
| SDSSJ135246.37+423923.5 | 2.30 | -28.14 | 9025.2 | 2.31 ±2.30 | <9.21 | <24.27 |
| SDSSJ142703.64+270940.3 | 1.17 | −26.08 | 442.4* | 4.726 ± 2.62 | <12.60 | <37.37 |
| SDSSJ155633.77+351757.3 | 1.50 | −26.86 | 15144.3 | 1.02 ± 2.23 | <7.71 | <55.52 |
| SDSSJ210712.77+005439.4 | 0.92 | −23.31 | 128.0 | −2.32 ± 2.77 | <5.60 | <54.16 |
| SDSSJ221511.93-004549.9 | 1.48 | −28.51 | 551.0 | −4.30 ± 2.84 | <4.22 | <49.58 |
| SDSSJ233646.20-010732.6 | 1.29 | −25.24 | − | −2.58 ± 2.84 | <5.60 | <50.57 |

**Table 2.** FeLoBAL QSO sample details and SCUBA-2 measurements. The absolute B-band magnitudes are derived from the absolute *i*-band (SDSS PSF) magnitude using a colour correction of $B-i = 0.35$ (Schneider et al. 2002). The balnicity index is taken from Gibson et al. (2009) if otherwise specified (* BI from Allen et al. 2011). Upper limits are computed by adding the $3\sigma$ value to the measured flux density. Due to the poor quality of the 450 $\mu$m data we only report the $3\sigma$ flux upper limits derived from these observations.

## 3 ANALYSIS AND RESULTS

### 3.1 850 $\mu$m flux density constraints of FeLoBAL QSOs

We measure the sub-mm flux densities and the relative errors at each SDSS optical position of the FeLoBALs in the 850 $\mu$m beam-convolved map and the noise map, respectively (Table 2). We achieved the depth we aimed for, i.e., $\sigma \simeq 2$ mJy at 850 $\mu$m ($1.9 \leqslant \sigma \leqslant 3.95$ mJy), however, none of the 17 FeLoBAL QSOs were individually detected at the $3\sigma$ level or above. Our measurements can provide useful 850 $\mu$m upper limits to help characterize the far-IR emission of FeLoBALs by constraining the Raileigh-Jeans tail of the dust emission to derive some crucial quantities such as IR luminosity, SFR and dust mass, which we discuss in Section 3.3 and 3.4.

Following previous studies of faint extra-galactic sources (e.g. Cao Orjales et al. 2012), we stacked the 850 $\mu$m maps in order to obtain higher S/N information on the average submm emission of thr FeLoBAL QSO population. A 40"×40" cutout of each submm map centred around each FeLoBAL SDSS position is created. Subsequently these cutouts are co-added via a weighted mean. The resulting stacked flux is extracted by simply reading out the value of the central pixel of the final image. This procedure can be summarized by the following mathematical expression:

$$F_s = \Sigma_{i=1}^{n} \frac{(F_i/\sigma_i^2)}{\Sigma_{i=1}^{n} 1/\sigma_i^2} \quad (2)$$

where $F_s$ is the stacked flux density, $F_i$ and $\sigma_i$ are the flux density and the noise corresponding to each source. The error on the stacked flux density is the inverse of the square root of the sum of all the inverse variances $1/\sigma_i^2$. For our sample of FeLoBALs we get $F_s = 1.14 \pm 0.58$ mJy (S/N~1.5).

| Source | SDSS flux (mJy) | 2MASS fluxes (mJy) | | | WISE fluxes (mJy) | | | | MIPS fluxes (mJy) | | |
|---|---|---|---|---|---|---|---|---|---|---|---|
| | 0.9$\mu$m | 1.2$\mu$m | 1.6$\mu$m | 2.2$\mu$m | 3.4$\mu$m | 4.6$\mu$m | 12$\mu$m | 22$\mu$m | 24$\mu$m | 70$\mu$m | 160$\mu$m |
| SDSSJ011117.34+142653.6 | 0.40±0.04 | 0.60±0.06 | 0.63±0.06 | 0.81±0.08 | 1.35±0.14 | 2.39±0.24 | 6.50±0.95 | 14.34±1.60 | 16.92±0.85 | 26.40±7.90 | 15.40±17.13 |
| SDSSJ024254.66-072205.6 | 0.08±0.01 | - | - | - | 0.55±0.08 | 1.04±0.15 | 2.15±0.20 | 2.16±0.50 | 6.40±0.64 | 10.80±4.50 | 16.20±14.16 |
| SDSSJ030000.57+004828.0 | 1.20±0.16 | 1.46±0.25 | 1.49±0.25 | 1.52±0.25 | 2.62±0.28 | 4.74±0.50 | 14.90±1.80 | 30.90±3.74 | 29.39±2.94 | 56.31±10.63 | -8.83±16.00 |
| SDSSJ083817.00+29526.5 | - | 0.43±0.04 | 0.58±0.05 | 0.60±0.10 | 0.46±0.01 | 0.53±0.02 | 2.22±0.16 | 6.46±0.16 | - | - | - |
| SDSSJ101108.89+515553.8 | - | 0.40±0.04 | 0.42±0.06 | 0.53±0.06 | 0.36±0.01 | 0.50±0.16 | 2.33±0.17 | 3.68±1.00 | - | - | - |
| SDSSJ102850.31+511053.1 | - | 0.44±0.05 | 0.60±0.07 | 0.91±0.06 | 0.46±0.02 | 0.44±0.02 | 1.22±0.13 | 1.88±1.20 | - | - | - |
| SDSSJ113424.64+323802.4 | - | 0.95±0.06 | 1.18±0.09 | 2.60±0.08 | 0.10±0.03 | 0.06±0.03 | 0.49±0.18 | 1.89±1.20 | - | - | - |
| SDSSJ113734.06+024159.3 | - | - | - | - | 0.31±0.01 | 0.37±0.01 | 1.55±0.15 | 3.28±1.20 | - | - | - |
| SDSSJ114509.73+534158.1 | - | 0.22±0.05 | 0.40±0.01 | 0.50±0.08 | 0.47±0.02 | 0.55±0.03 | 1.97±0.13 | 4.03±1.00 | - | - | - |
| SDSSJ123549.95+013252.6 | 0.11±0.02 | 0.28±0.05 | 0.22±0.05 | - | 0.30±0.04 | 0.47±0.07 | 1.34±0.30 | 3.00±0.30 | 4.70±0.30 | 30.00±7.50 | 22.70±10.80 |
| SDSSJ131957.70+283311.1 | - | 0.85±0.05 | 0.81±0.07 | 1.24±0.07 | 1.05±0.01 | 1.27±0.04 | 3.99±0.16 | 6.80±1.06 | - | - | - |
| SDSSJ135246.37+423923.5 | - | 1.18±0.05 | 1.29±0.08 | 1.90±0.07 | 1.34±0.03 | 1.59±0.04 | 5.50±0.16 | 11.4±0.90 | - | - | - |
| SDSSJ142703.64+270940.3 | 0.22±0.02 | - | - | - | 0.57±0.06 | 1.15±0.12 | 3.05±0.31 | 6.76±0.66 | 5.60±0.56 | 36.80±6.30 | 69.00±14.37 |
| SDSSJ155633.77+351757.3 | 0.69±0.10 | 1.10±0.20 | 0.81±0.16 | - | 1.64±0.19 | 3.52±0.50 | 10.01±1.50 | 17.96±2.50 | 16.50±1.65 | 31.50±5.08 | 35.40±18.61 |
| SDSSJ210712.77+005439.4 | 0.04±0.01 | - | - | - | 1.56±0.16 | 3.93±0.40 | 12.04±1.80 | 24.34±4.00 | 20.10±2.10 | 44.10±8.16 | 60.30±20.80 |
| SDSSJ221511.93-004549.9 | 0.87±0.09 | 1.13±0.15 | 0.88±0.10 | - | - | - | - | - | 12.20±0.65 | 33.60±5.79 | 42.10±21.03 |
| SDSSJ233646.20-010732.6 | 0.12±0.02 | 0.20±0.03 | 0.21±0.03 | - | - | - | - | - | 0.88±0.14 | 2.19±6.73 | 20.19±16.34 |

**Table 3.** Multiwavelength photometry of the 17 FeLoBALs. SDSS fluxes are taken from SDSS DR6. NIR photmetry from the public 2MASS All Sky Point Source Catalog. WISE fluxes from the operational database as of 2011 June (Cutri et al. 2012) and MIPS data from Farrah et al. (2012).

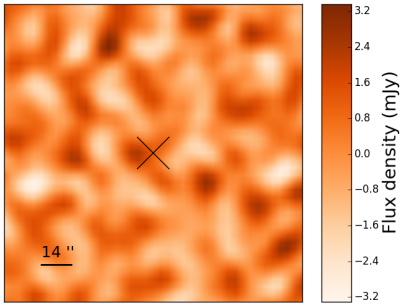

**Figure 2.** 144 x 144 arcsec$^2$ SCUBA-2 850 $\mu$m stacked map of the 17 FeLoBAL QSOs of our sample (the SCUBA-2 beam size is 15 arcsec). The cross at the centre of the image marks the position of the stack. The colour scale in mJy goes from blue to red, with red areas indicating higher flux. The stacking procedure used to produce this image is fully described in Section 3.1.

### 3.2 Statistical Analysis

The most direct way to test whether or not FeLoBALs represent a highly star-forming stage in the life of young QSOs is through a comparison of their sub-mm properties with other classes of QSOs. We consider four different samples: (1) the FeLoBALs of this paper; (2) the BAL QSOs of Willott et al (2003); (3) the BAL QSOs of Priddey et al (2007); (4) and the non-BAL QSOs of Priddey et al (2003). Willott et al. (2003) observed 30 BAL quasars with $2 \leqslant z \leqslant 2.6$ using the SCUBA camera at the JCMT. This sample was drawn from the SDSS Early Data Release BAL sample of Reichard et al. (2003). Priddey et al. (2003) performed SCUBA observations of 57 non-BAL quasars from the Large Bright Quasar Survey (LBQS; Hewett et al. 1995) in the redshift range 1.5 $\leqslant z \leqslant 3$; and Priddey et al. (2007) observed 15 LBQS BAL QSOs with SCUBA.

Willott et al. (2003) included in their control sample also a list of 35 normal quasars from Omont et al. (2003) observed at 1.2 mm with the Max-Planck Millimeter Bolometer Array (MAMBO) at the Institut de Radioastronomie Millimetrique (IRAM) 30-meter telescope, which we do not include here in order to avoid any errors introduced by the conversion factor between flux measurements at different sub-mm wavelengths.

As shown in Figure 1, our sample of FeLoBALs has a slightly different redshift and magnitude distributions to the benchmark samples. Our sample redshift varies between 0.89 and 2.78, while the comparison samples span a less extensive range, with $1.8 \leqslant z \leqslant 2.9$. The median redshift of the FeLoBAL, BAL and non-BAL samples are z=2.04, z=2.12, and z=2.23, respectively. These differences are also confirmed by the results of Kolmogorov-Smirnov tests, which returns 95% (97.5%) probability that FeLoBALs and BALs (non-BALs) are not drawn from the same population in terms of redshift. The dissimilarity in redshift could represent a caveat to our study, however this difference can be neglected if it can be plausibly assumed that the mechanism producing BALs does not depend strongly in redshift.

Regarding the differences in the optical absolute magnitudes, the FeLoBALs of this paper as well as the BALs from Willott et al. (2003) were selected from the SDSS, and are thus less luminous than the LBQS quasars of Priddey et al. (2003) and Priddey et al. (2007) (LBQS is a catalog of QSOs comparable in brightness with the high luminous quasars at z=4). The median B-band absolute magnitude of the FeLoBALs, BALs and non-BALs samples are $M_B$=−27.16, $M_B$=−27.46 and $M_B$=−28.3, respectively. Much effort in the past was dedicated to the study of a link between optical and sub-mm luminosities (Omont et al. 2003; Willott et al. 2003). Even though a weak connection between these two quantities seems to exist, no statistically significant correlation was found. For this reason we might expect the differences in magnitude between our sample and the benchmark samples to affect the reliability of our results. For each of these samples we calculated the 850 $\mu$m flux density straight mean and weighted mean with the respective errors. We find that the simple mean and the weighted mean of the flux densities of FeLoBALs are consistent within $2\sigma$ of the error bars with those of both BALs and non-BALs, even though they appear to be slightly lower. For this reason it is worth investigating the possbile effect on this result caused by the difference in the magnitude distribution of the samples. If we stack the fluxes of the FeLoBALs brighter than $M_B = -26$, the value we obtain is $F_{850} = 1.38 \pm 0.74$ mJy. On the contrary, the sub-sample made up of the fainter FeLoBALs ($M_B \geqslant -26$) is characterized by a stacked flux of $F_{850} = 0.45 \pm 1.02$ mJy; the values obtained are consistent within $1\sigma$. As proposed by previous studies, our results also show the lack of a significant connection between absolute magnitude and sub-mm flux.

| Sample | QSO type | Median z | Median $M_B$ | N | Mean $F_{850}$ (mJy) | Weighted mean $F_{850}$ (mJy) |
|---|---|---|---|---|---|---|
| This paper | FeLoBAL | 2.04 | -27.16 | 17 | $0.83 \pm 0.68$ | $1.14 \pm 0.58$ |
| P03 | Non-BAL | 2.23 | -28.30 | 25 | $2.88 \pm 0.79$ | $2.83 \pm 0.55$ |
| P07 | BAL | 2.13 | -28.00 | 16 | $2.76 \pm 0.79$ | $2.38 \pm 0.32$ |
| W03 | BAL | 2.16 | -27.20 | 30 | $2.56 \pm 0.67$ | $2.55 \pm 0.45$ |
| P07+W03 | BAL | 2.12 | -27.46 | 46 | $2.63 \pm 0.60$ | $2.44 \pm 0.26$ |

**Table 4.** Sub-mm properties of our FeLoBAL sample and comparison samples. References: P03 – Priddey et al. (2003); P07– Priddey et al. (2007); W03 – Willott et al. (2003). $N$ is the number of sources in each sample. The last column is the weighted mean with the associated error obtained following the procedure described in section 3.1.

We also test if the 850 $\mu$m flux density distribution of FeLoBALs is consistent with the distributions of BALs and non-BALs. Due to the lack of significant detections in the samples we performed a survival analysis test which can account for the presence of upper limits (Gehan test; Isobe & Feigelson 1986). This non-parametric test analyses the difference in the sample distributions and returns the probability that this difference occurs by chance. Comparing FeLoBALs and BALs the result obtained is 22%; we cannot therefore reject the null hypothesis that FeLoBAL and BAL quasars are drawn from the same 850 $\mu$m flux density distribution. Between FeLoBALs and non-BALs the probability is instead 5%. This result may suggest a hint of discrepancy in the two distributions, however, it is still not significant enough and again we cannot reject the null hypothesis.

In conclusion, the 850 $\mu$m emission of FeLoBALs does not appear to be different from that of other types of QSOs. In particular, our analyses show no indication of higher energy output in the submm flux from FeLoBALs over other types of QSOs, as would be expected if FeLoBALs are an intermediate stage between starburst and an obscured QSOs.

*3.2.1 Correlation between sub-mm emission and balnicity*

An issue that could affect our statistical results is the different methods chosen in the selection of BAL QSOs in each sample. For instance, Willott et al. (2003) rejected objects with very weak outflows ( BI $\leqslant$ 200 km s$^{-1}$). For our study we did not apply any restriction based on the BI values and our sources span the range 0 km s$^{-1}$ $\leqslant$ BI $\leqslant$ 18000 km s$^{-1}$; therefore it is essential to check whether this difference could potentially affect our comparison of the sub-mm properties. Willott et al. (2003) found no correlation between the 850 $\mu$m flux density and BI (although the species they selected to calculate the balnicity is the CIV), while Priddey et al. (2007) found a tentative positive correlation of sub-mm flux density with the equivalent width (EW) of CIV absorption together with a link between 850 $\mu$m detection rate (at level $\geqslant 2\sigma$) and EW. This relation however becomes less significant once BI is taken into account instead of the EW.

Due to the lack of $2\sigma$ detections in our sample any study on the correlation between FeLoBALs sub-mm flux and balnicity would not provide any meaningful information. We can however split the sample into two bins to separate sources with weaker BALs (BI $\leqslant$ 1000 km s$^{-1}$) from those with stronger ones (BI $\geqslant$ 1000 km s$^{-1}$), the flux density weighted means of the two sub-samples are respectively $0.82 \pm 0.91$ mJy and $1.55 \pm 1.08$ mJy, which are consistent within $1\sigma$. In conclusion we do not find any evidence to support a link between sub-mm flux density and BI and our result seems to agree with those of previous studies, although any comparison between our analysis and previous works must be taken with extra care as the species used to derive the BI differ.

### 3.3 SED fits

We performed individual SED fitting from the near-IR up to the submm wavelengths in order to characterize the emission of FeLoBALs. We use data from the Sloan Digital Sky Survey Data Release 6 (SDSS DR6, York et al. 2000); the Two Micron All-Sky Survey (2MASS, Skrutskie et al. 2006), the Wide-Field Infrared Survey Explorer (WISE, Cutri et al. 2012; Wright et al. 2010; Jarrett et al. 2011) and from the Multiband Imaging Photometer for *Spitzer* (MIPS) (Rieke et al. 2004; Farrah et al. 2012).

We initially use six empirical templates from the Spitzer Wide-area InfraRed Extragalactic (SWIRE) library of Polletta et al. (2006): the SEDs of the starburst galaxies M82 and Arp220, the star forming QSO Mrk231, and three SEDs derived by combining spectra, models and photometric data which reproduce two type 1 and one type 2 quasars (which therefore also include a contribution in the energy output from star formation). For all 17 sources in our sample the best SED fit is achieved by using the quasar models, furthermore the far-IR photometry does not indicate any excess over the templates. On the contrary, the starburst galaxy templates well overestimate the observed far-IR emission. This suggests that FeLoBAL SEDs do not differ much from that of normal quasars and that they have similar star formation activity. The best fit $\chi^2_{red}$ values from the fits vary between 3.4 0and 119.10 (see Figure 3 and 4). The poor qual-ity of the fits is mostly due to the difficulty of the models in reproducing accurately the emission in the NIR (2MASS data) which is underestimated once the far-IR photometry is matched.

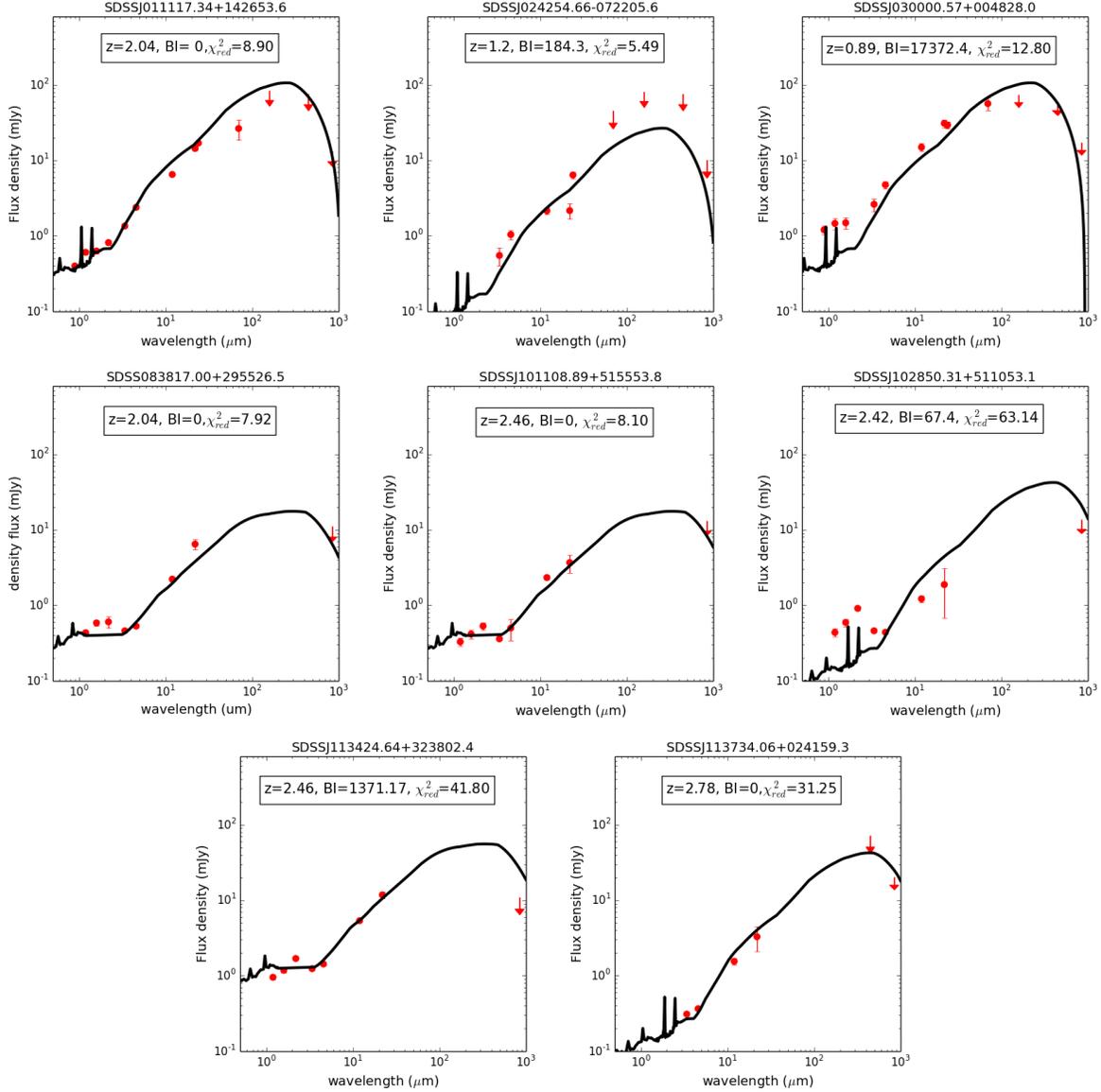

**Figure 3.** Observed frame optical to sub-mm FeLoBAL SEDs single template best fits. The black line (and also single black points in the near-IR) represents an empirical template from the SWIRE library (Polletta et al. 2006). The models are described in Section 3.3. The best fitting $\chi^2_{red}$ values for each object are given in Table 4. For all the 17 FeLoBALs of our sample the best fit is achieved by the use of a QSO template.

To remedy this issue we proceed with a more complex method and we consider three different components to reproduce the SEDs: stellar, AGN and starburst emission. In particular, we are interested in disentangling the star formation emission from the AGN emission, and also understanding whether or not the sub-mm emission can be simply described by nuclear activity or instead if a starburst component is needed. For the stellar emission we use a library of stellar population templates by Bruzual et al. (2003) which has been shown to reproduce the continuum and line emission of galaxies in the SDSS catalogue well (Tremonti et al. 2003). Each template varies both in metallicity (Z=0.08, Z=0.2, Z=0.5) and stellar age (25 Myr, 100 Myr, 290 Myr, 640 Myr, 900 Myr, 1.4 Gyr, 2.5 Gyr, 5 Gyr, 6 Gyr, 12 Gyr). For the AGN and starburst components we rely on a set of templates by Efstathiou & Rowan-Robinson (1999) and Efstathiou & Siebenmorgen (2009). AGN emission varies with viewing angle and the model assumes a smooth torus whose density and thickness are, respectively, inversely and directly proportional to the distance from the nucleus (the use of smooth torus models may represent a caveat since recent studies provided observational evidence in support of a clumpy morphology, e.g. Mullaney et al. 2011). The starburst models have 15 different starburst ages in the range 0-70 Myr, equally spaced by 5 Myr, and each of these was produced according to three different values of $K\tau$ (50, 100, 150), where $e^{K\tau}$ represents the attenuation. These starburst models also take into account the effects of the presence of PAHs. Again we consider all the photometry available for each source and fluxes with detections $\leqslant 3\sigma$ are included as normal values with their error bars. In this way our study is not affected by the choice of the level of significance of the measurements. In the fitting procedure all of the possible combinations of the three components (stellar, AGN and starburst) are included, which means that we do not make any a priori assumptions about the composition of the SED. The best fit SEDs are presented in Figures 5, 6 and 7, with the best fitting $\chi^2_{red}$ values ranging from 0.8 to 18.24. Strictly speaking, for two sources, SDSSJ142703.64+270940.3 and SDSSJ024254.66-072205.6, the new fit is worse in a statistical $\chi^2$ sense than the one performed with a single template from the Polletta library, but for the rest of the sample there is a remarkable improvement. Some sources have $\chi^2_{red} \geqslant 3$, however this is predominantly due to difficulties reproducing the near-IR photometry; the far IR SEDs are generally reproduced well. For 6/17 FeLoBALs (SDSSJ024254.66-072205.6, SDSSJ123549.95+013252.6, SDSSJ142703.64+270940.3, SDSSJ155633.77+351757.3, SDSSJ210712.77+005439.4 and SDSSJ221511.93-004549.9) the inclusion of a starburst component improves the SED fit, whereas for the rest of the sample the far-IR emission is well-described by an AGN template and an additional starburst component is unnecessary. In general, an enhanced activity in the far-IR powered by star formation does not seem to be a common feature for the majority of FeLoBALs. This result is in agreement with indication provided by the previous fits, which showed that FeLoBAL QSOs have far-IR emission similar to those of regular quasars. We discuss this point further in Section 4.

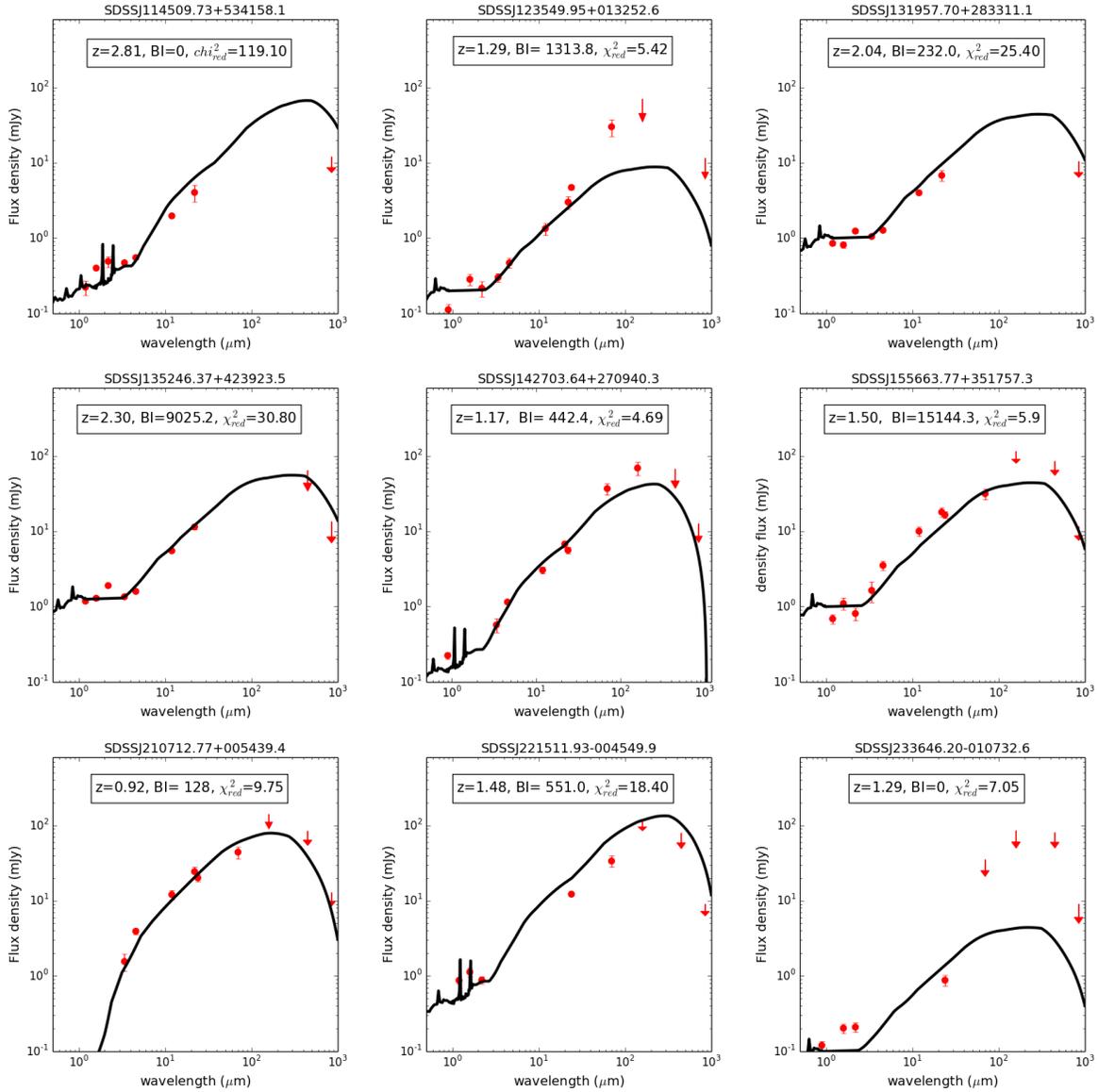

**Figure 4.** Observed frame optical to sub-mm FeLoBALs SEDs single template best fits. Details as in Figure 3.

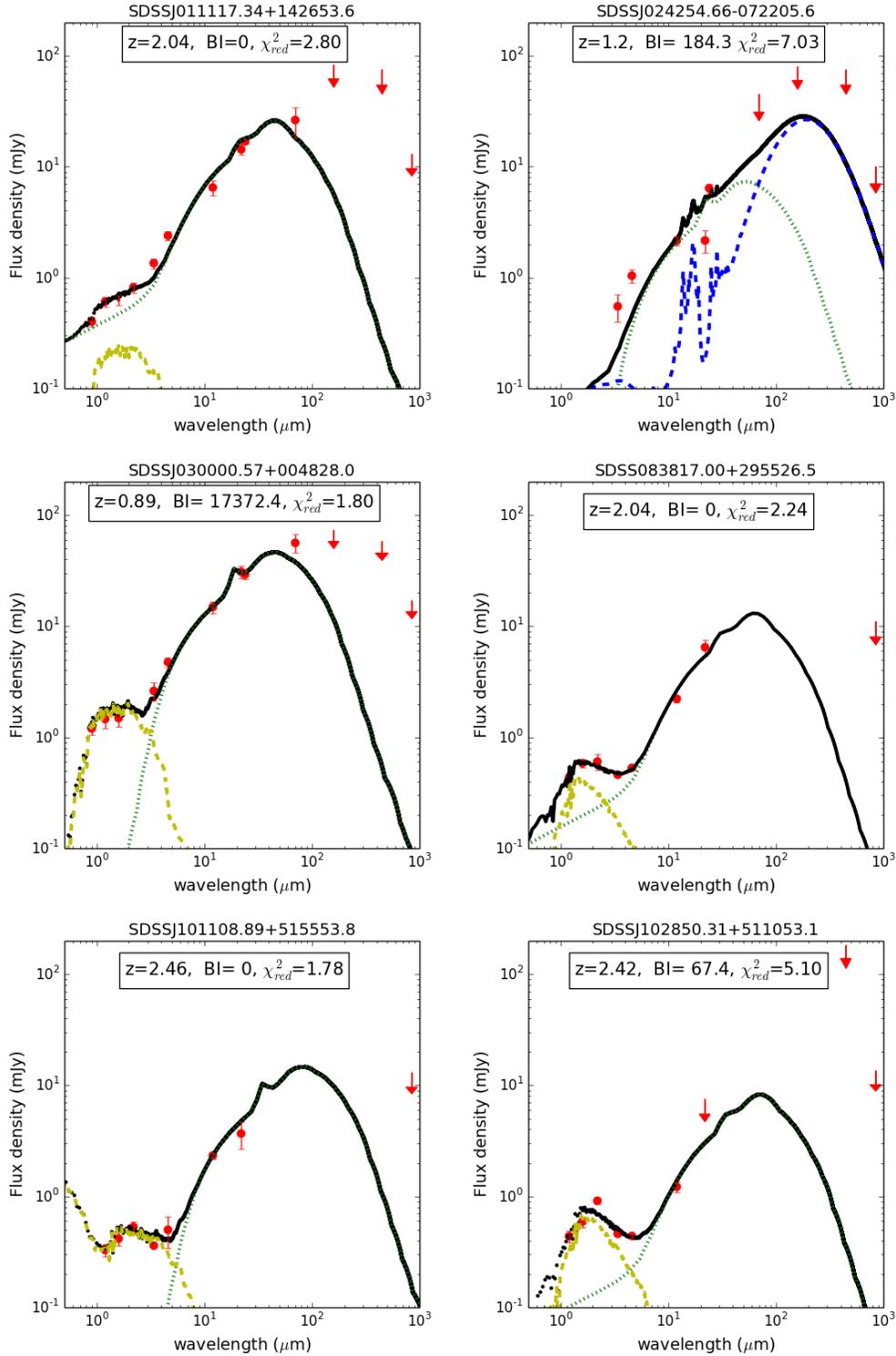

**Figure 5.** Observed-frame optical to sub-mm SED best fits of FeloBALs, using the approach described in Section 3.3. The yellow line is the stellar component, the cyan is the AGN component, the blue line is the starburst component and the black line is the best fit composite model. Measurements with $S/N \leqslant 3$ are plotted as upper limits. The Best fit reduced chi-square values for each source are listed in Table 4, a description of the different model is presented in Section 3.3. The inclusion of a starburst component improve the fit only in 6/17 objects, for the remaining sources the IR emission is consistent with being dominated by AGN activity.

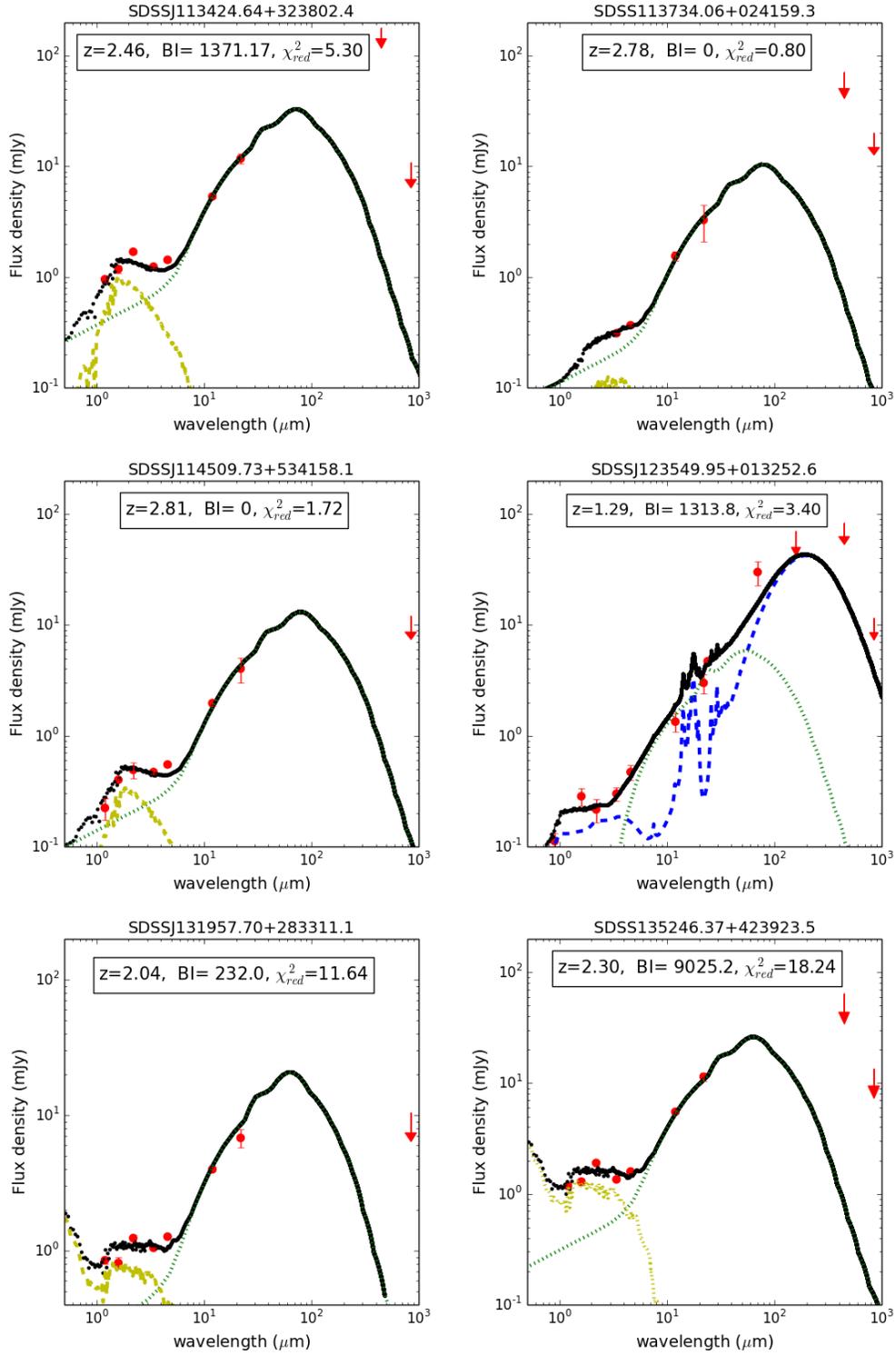

**Figure 6.** Observed-frame optical to sub-mm SED best fits of FeloBALs, details are the same of Fig. 5.

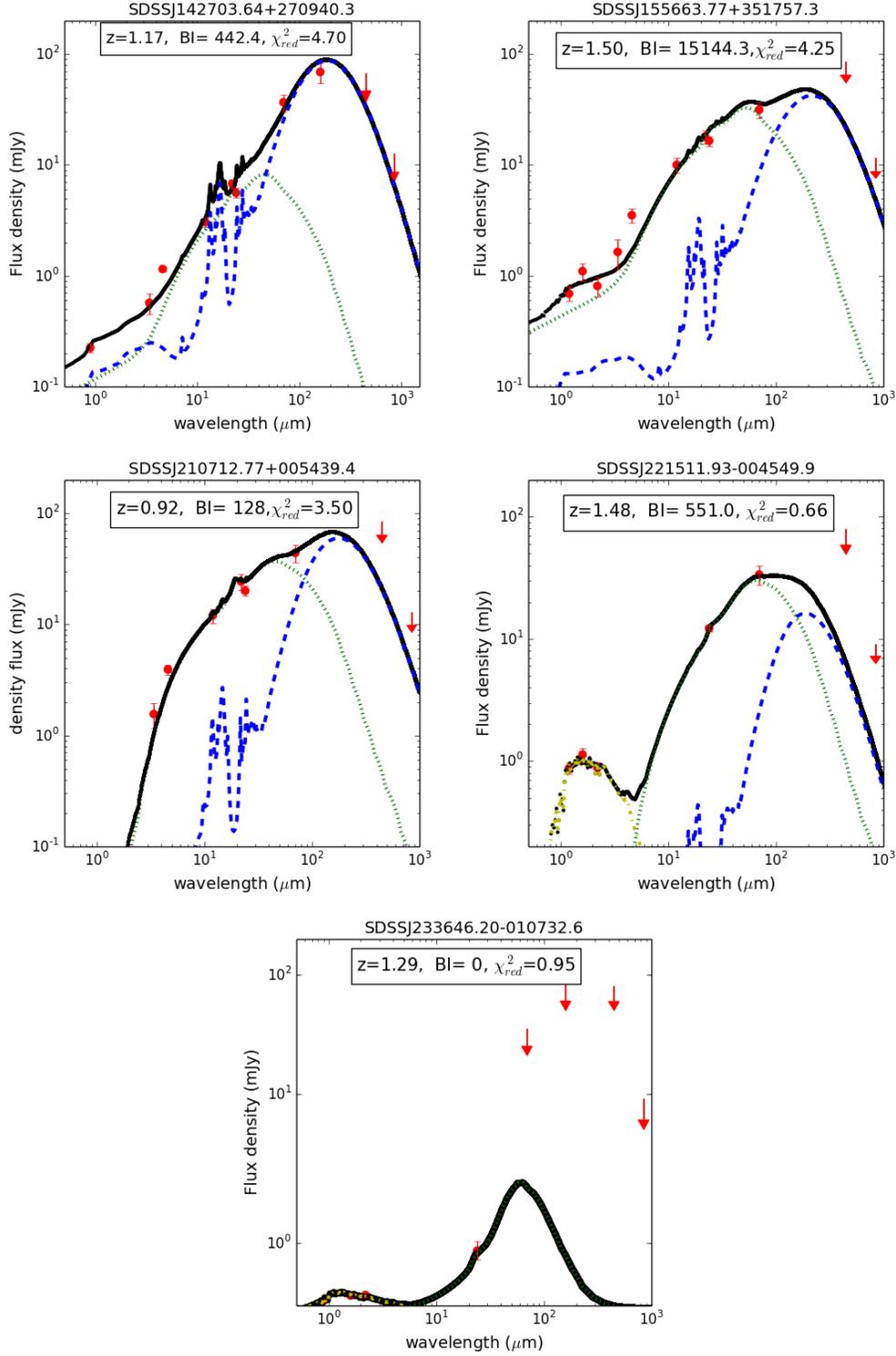

**Figure 7.** Observed-frame optical to sub-mm SED best fits of FeLoBALs, details are the same of Fig. 5.

### 3.3.1 Infrared luminosities and star-formation rates

We determine IR luminosities by integrating under the best-fitting SED of each source over the rest-frame wavelength range 8-1000 $\mu$m (see Table 5). All but one of the objects have $L_{IR} \geqslant 10^{12}$ $L_\odot$, large enough to be classified as ULIRGs. The only exception is SDSSJ233646.20-010732.6, whose IR luminosity is of the order of $10^8$ $L_\odot$; which explains the complete lack of detection of this source in the MIPS bands. To evaluate the errors on the IR luminosities we proceed as follows: for each combination of SED components, we record the best-fit $\chi^2$ and use these values to generate relative probabilities for each combination by assuming that $P \propto \exp(-\frac{\chi^2}{2})$. We then marginalise these values and use them to generate a cumulative frequency distribution (CFD) of far-IR luminosity for each object. Errors on the best-fit luminosity (which is derived using the model combination that minimises $\chi^2$) are quoted in table 5 and correspond to half the difference between the 84th and 16th percentiles of the CFD; they are equivalent to $1\sigma$ uncertainties in the limit of Gaussian uncertainties. The IR luminosity is consistent with being powered by an AGN for 11/17 FeLoBALs, while for the remaining 6 objects a combination of AGN and star-formation appear to be responsible.

To evaluate the star-formation rates of our FeLoBALs we employ the classic Kennicutt conversion (Kennicutt 1998):

$$SFR(M_\odot \text{yr}^{-1}) = 4.5 \times 10^{-44} \, L_{IR} \, (\text{erg s}^{-1}) \quad (3)$$

The luminosity that appears in this formula strictly represents the thermal emission of dust which reprocesses the absorbed optical and UV radiation field emitted by young O and B stars in the star-forming region. As a consequence, any contribution to the IR luminosity arising from the dust heated by other sources (AGN, old stars) is not accounted for. We take a conservative approach and therefore we estimate the SFR only for those sources whose SED best fitting includes a SB template. The $L_{IR}$ that appears in equation 3 is simply calculated by integrating under the best fitting SB model. SFRs are of the order of $10^2 - 10^3$ $M_\odot$ yr$^{-1}$ (Table 4), and these values are consistent within errors with those in both Farrah et al. (2010) and Farrah et al. (2012), where FeLoBALs SFRs were calculated using the Polycyclic Aromatic Hydrocarbon (PAH) luminosities and the monochromatic luminosities at 60 $\mu$m, respectively, which yields SFR $\sim 10^2$-$10^3$ $M_\odot$ yr$^{-1}$.

### 3.4 ISM mass estimates

The 850 $\mu$m flux density measurements can also be used to constrain the dust masses in the host galaxies of the FeLoBAL quasars. The assumption here is that the sub-mm flux traces the optically thin thermal emission of dust which reprocesses both the UV–optical light from young O-B stars and the AGN radiation. The total dust mass can be written as:

$$M_{\text{dust}} = \frac{1}{1+z} \frac{S_{850} D_L^2}{k_d^{\text{rest}} B(\nu^{\text{rest}}, T)} \quad (4)$$

where $S_{850}$ is the flux density at 850 $\mu$m, $D_L$ is the luminosity distance, $k_d^{\text{rest}}$ is the rest-frequency mass absorption coefficient and $B(\nu^{\text{rest}}, T)$ is the rest-frequency Planck function at temperature $T$. The value of $k_d^{\text{rest}}$ at 850 $\mu$m varies between 0.04 and 0.3 m$^2$ Kg$^{-1}$ (Mathis & Whiffen, 1989), and we use an intermediate value of 0.15 m$^2$Kg$^{-1}$ as in Chini et al. (1986). For the dust temperature we assume $T$=35 K, which is a typical value of nearby starburst galaxies (Scoville et al. 2014). Since the FeLoBALs were not significantly detected, we use the $S_{850}$ $3\sigma$ upper limits and therefore the values of the masses must be considered as upper limits as well (see Table 4). These results are affected by the uncertainty on the temperature $T$, although the results do not dramatically change as long as the dust temperature falls in the range 20 - 45 K (Hughes et al. 1997). The dust mass upper limits we calculated are consistent with the typical values of dust-rich systems such as sub-millimetre galaxies (SMG), whose dust content is of the order of $M_{\text{dust}} = 9 \times 10^9 M_\odot$ (e.g, Toft et al. 2014; Kovacs et al 2006).

## 4 DISCUSSION

We have tested the idea that FeLoBAL QSOs represent a transition phase between a young, dusty, starburst quasar and an optically luminous quasar where star formation is being quenched by AGN feedback. Specifically, our attention has been focused on the submm emission of FeLoBAL QSOs in order to look for evidence of enhanced star formation activity. An alterntive picture describes FeLoBALs as normal QSOs whose special features can be described simply by invoking an orientation effect (e.g. Elvis et al. 2000).

The nature of BAL QSOs has been the subject of numerous far-IR investigations in the past and we briefly consider these in light of our new submm observations. Willott et al. (2003) carried out a statistical analysis based on SCUBA observations of 57 BAL QSOs. By comparing the sub-mm properties of his objects with a composite sample of non-BALs from Priddey et al. (2003) and Omont et al. (2003) he showed that BALs are statistically undistinguishable from normal quasars. Priddey et al. (2007) performed a similar investigation and reached the same conclusion, i.e. BALs are not brighter sources in the submillimetre. Since the samples employed in these studies were mostly composed of HiBALs, our study based exclusively on FeLoBALs represents a completion of these previous investigations, and confirms that the population of BAL quasars as a whole is not characterized by higher sub-mm fluxes ($L_{FIR}=10^{13}$ $L_\odot$), as expected if they represent the termination of a starburst galaxy. The 850 $\mu$m flux density weighted mean of our sample of FeLoBALs is $F_s = 1.14 \pm 0.58$ mJy. By applying the Kennicutt SFR-$L_{IR}$ conversion (Kennicutt 1998) and assuming a variety of dust templates (Chary & Elbaz 2001, Efstathiou & Siebenmorgen 2009) this value corresponds to a SFR of $\sim$ 150–240 $M_{\odot}yr^{-1}$. This result suggests that FeLoBAL QSOs are forming new stars at similar rates of both BAL and non-BAL quasars ($\sim 10^2 M_{\odot}yr^{-1}$, Cao Orjales et al. 2012), and not at the prodigious rates typical of luminous starburst galaxies, such as SMGs, which are characterized by SFRs$\geqslant$500 $M_{\odot}yr^{-1}$ (e.g. Magnelli et al. 2012).

| Source | $log(M_{dust}/M_{\odot})$ | $L_{ir}^{tot}$ ($10^{12}$ $L_{\odot}$) | SFR ($M_{\odot}yr^{-1}$) | $\chi^2_{red}$(Single) | $\chi^2_{red}$(Composite) |
|---|---|---|---|---|---|
| SDSSJ011117.34+142653.6 | <9.3 | 7.60±0.38 | - | 8.90 | 2.80 |
| SDSSJ024254.66-072205.6 | <8.7 | 4.28±1.17 | 331±141 | 5.49 | 7.03 |
| SDSSJ030000.57+004828.0 | <8.5 | 15.60±0.50 | - | 12.80 | 1.80 |
| SDSSJ083817.00+295526.5 | <9.3 | 2.70±0.17 | - | 7.92 | 2.24 |
| SDSSJ101108.89+515553.8 | <9.2 | 2.70±0.44 | - | 8.10 | 1.78 |
| SDSSJ102850.31+511053.1 | <9.3 | 1.50±0.80 | - | 63.14 | 5.10 |
| SDSSJ113424.64+323802.4 | <9.2 | 5.90±0.42 | - | 41.80 | 5.30 |
| SDSSJ1137340.6+024159.3 | <9.6 | 1.74±0.19 | - | 31.25 | 0.80 |
| SDSSJ114509.73+534158.1 | <9.3 | 2.13±0.11 | - | 119.10 | 1.72 |
| SDSSJ123549.95 +013252.6 | <8.8 | 7.04±0.60 | 703±70 | 3.40 | 2.60 |
| SDSSJ131957.70+283311.1 | <9.1 | 3.40±0.35 | - | 25.40 | 11.64 |
| SDSSJ135246.37+423923.5 | <9.1 | 5.34±1.18 | - | 30.80 | 18.24 |
| SDSSJ142703.64+270940.3 | <8.7 | 5.44±0.21 | 610±74 | 4.69 | 4.70 |
| SDSSJ155633.77+351757.3 | <8.9 | 12.50±3.00 | 794±75 | 5.90 | 4.25 |
| SDSSJ210712.77+005439.4 | <8.5 | 14.25±2.00 | 348±11 | 9.75 | 3.50 |
| SDSSJ221511.93-004549.9 | <8.9 | 8.15±0.45 | 254±120 | 18.40 | 0.66 |
| SDSSJ233646.20-010732.6 | <8.8 | 0.0004±0.0001 | - | 7.05 | 0.95 |

**Table 5.** Dust mass upper limits, IR luminosities and SFRs of FeLoBAL QSOs calculated in Section 3.4. In the fifth and in the sixth columns we report the best fitting $\chi^2_{red}$ values derived from the single Polletta template fit and the composite (AGN, starburst and stellar components) fit, respectively. $1\sigma$ errors on IR luminosities and SFRs are quoted.

The FeLoBALs SEDs fitting analysis may however suggest another scenario which does not completely rule out the hypothesis that some FeLoBALs have enhanced starformation, which we now discuss. For 6/17 FeLoBALs a starburst component is needed to best describe the emission in the FIR. All of these sources have MIPS observations and are characterized by SFRs, in the range 250–800 $M_{\odot}yr^{-1}$(hence higher than 'normal' QSOs). The SEDs analysis is of course affected by the choice of the stellar, AGN and starburst templates, notwithstanding, the models we used successfully reproduce the emission of local star forming galaxies and AGNs (Ruiz et al. 2001; Farrah et al. 2002; Verma et al. 2009), therefore the use of different models would give the same overall conclusions. This result may indicates that some FeLoBALs are undergoing a 'hot' starburst phase, in which the burst of star formation heats up the dust grains up to T$\geqslant$70 K. At this temperature the dust thermal black body emission peaks at shorter wavelengths, and therefore would not leave an evident trace in the submm (Acosta-Pulido et al. 1999; Klaas et al. 1999). In this scenario some FeLoBALs could in principle still constitute a transition phase between a starburst galaxy and unobscured quasar despite the non detection in the submm.

Another peculiarity of FeLoBALs which deserves particular notice is the presence of AGN-driven outflows which, extending up to kpc scale, are thought to have a crucial impact on the host galaxy (de Kool et al. 2002; Cicone et al. 2014; Harrison et al. 2014). In a recent study conducted by Farrah et al. (2012) on FeLoBAL QSOs a clear anticorrelation between the starburst contribution to the IR luminosity and the BI is shown. This may reveal the disrupting effect of AGN feedback on star formation. In our study only 6/17 sources required a starburst component in the best-fitting SED,and as a consequence we could not look for a trend between outflow strength and emission from star formation. We do however note that the FeLoBALs which do display photometric evidence for the presence of a starburst, span a wide and varied range of balnicity, which may suggest that the host galaxy is not affected by the violent AGN-driven gas outflows which may take place on short distance scales, e.g. $10^{-2}$ pc (e.g. Capellupo et al. 2012).

In the past, studies have been conducted in different wavelength regimes with an aim to better understand the nature of BAL QSOs, yelding different conclusions. Becker et al. (2000) analyzed the radio properties of BAL QSOs and argue that the differences seen with respect to the morphologies and spectral indices compared to normal quasars cannot be explained by simply invoking an orientation model. Di Pompeo et al. (2013) found a statistically excess in mid-IR luminosities of radio-loud BAL QSOs. Moreover, studies based on NuSTAR observations showed that the hard X-ray emission (8–24 keV) of BALs tend to be intrinsically weaker than that of non-BALs (Teng et al. 2014, Luo et al. 2014), even though in a previous *Chandra* survey conducted by Gallagher et al. (2006) it was inferred that that BALs and non-BALs have the same X-ray fluxes in the energy range 0.8–8 keV once intrinsic absorption is taken into account. The same conclusion was reached by Weymann et al. (1991) noticing the similarity of the optical emission lines of BALs and other QSOs. However, due to the very small percentage of FeLoBALs in the whole population of optically selected QSOs (0.3%,Trump et al. 2006) most of the samples in the cited works are mainly made up of HiBALs. Thus an extrapolation of results based HiBALs to the FeLoBAL class of QSOs may be misleading.

Our study does not provide any clear indication that FeLoBAL QSOs are characterized by the presence of a lu-minous cold starburst, at least in the majority of cases. As a consequence, either the star formation rate in this class of quasars is significantly lower than typical starbursts, or it is present but unusually hot and therefore does not leave any trace in the sub-mm. These results argue against the hypothesis that FeLoBALs embody an intermediate phase between a ULIRG and an unobscured quasar, although this hypothesis cannot be completely ruled out.

For instance, Urrutia et al. (2009) found a large fraction of FeLoBAL QSOs which displayed reddened optical/UV spectra, indicating that they reside in heavily dust-enshrouded environments where increased star formation may occur. Furthermore, in a follow-up study based on Spitzer observations, Urrutia et al. (2012) pointed out that these type of sources lie below the BH mass–host luminosity relation and therefore they argued that these red QSOs could in principle still constitute an intermediate stage, in which the merger-induced starburst has occured long before the black hole began its growth.

In conclusion the evolutionary scenario drawn by Sanders et al. (1988a) still remains puzzling; more work must be done in order to understand whether or not FeLoBAL QSOs are transition objects, which stage of the transition they represent, and how they relate to other potential intermediate sources like Hot Dust-Obscured Galaxies (Hot DOGs, Assef et al. 2014) and WISE/radio-selected AGN (Jones et al. 2015).

## 5 CONCLUSION

In this paper we present the results derived from SCUBA-2 850$\mu$m observations of 17 FeLoBAL QSOs. These constitute the largest sample of this class of quasars ever observed at these wavelengths.
We concluded the following:
1) FeLoBAL QSOs are not exceptionally bright sources in the submillimetre. Statistical and survival analyses reveal that they have submm properties which are indistinguishable from those of BAL QSOs and normal quasars.
2) FeLoBALs have total IR luminosities of the order of $10^{12} L_\odot$ and can therefore be classified as ULIRGs. The long-wavelength SEDs of the majority of FeLoBALs are similar to those of normal QSOs. Our SED fitting analysis shows that the observed far-IR emission from most FeLoBALs is consistent with being dominated by AGN activity. For only 6/17 sources in our sample is the fit improved with the inclusion of a starburst component.
3) Our results indicate that FeLoBAL QSOs are not undergoing a 'cold' starburst phase and we do not find evidence suggesting that FeLoBALs universally represent an intermediate stage between a highly star- forming galaxy and young obscured QSO. The presence of an exceptional 'hot' starburst event cannot be completely ruled out even though such a component seems unlikely.

## 6 ACKNOWLEDGMENTS

GV acknowledges the University of Hertfordshire for a PhD studentship. KEKC acknowledges support from the UK Science and Technology Facilities Council [STFC; grant number ST/M001008/1, ST/J001333/1]. The Dark Cosmology Centre is funded by the Danish National Research Foundation. DMA and JLW acknowledge support from STFC [grant number ST/I001573/1]. JEG acknowledges support from the Royal Society. JLW is supported by a European Union COFUND/Durham Junior Research Fellowship (under EU grant, agreement number 267209). We thank the SCUBA-2 instrument and software teams, and the Joint Astronomy Centre staff for taking the data for our programmes M09BI104 and M12AC07. The James Clerk Maxwell Telescope has historically been operated by the Joint Astronomy Centre on behalf of the STFC of the United Kingdom, the National Research Council of Canada and the Netherlands Organisation for Scientific Research. Additional funds for the construction of SCUBA-2 were provided by the Canada Foundation for Innovation. This publication makes use of data products from the Two Micron All Sky Survey, which is a joint project of the University of Massachusetts and the Infrared Processing and Analysis Center/California Institute of Technology, funded by the National Aeronautics and Space Administration and the National Science Foundation.